\documentclass[floatfix,aps,prd,abstract,10pt,nofootinbib,preprintnumbers,twocolumn]{revtex4}

\usepackage{amsgen,amsmath,amstext,amsbsy,amsopn,amssymb}

\usepackage[dvips]{color}

\usepackage{graphicx}
\usepackage{epsfig}
\usepackage{multirow}
\usepackage{longtable}

\begin{document}

\newcommand{\beq}{\begin{equation}}
\newcommand{\eeq}{\end{equation}}
\newcommand{\beqa}{\begin{eqnarray}}
\newcommand{\eeqa}{\end{eqnarray}}

\def\la{\mathrel{\mathpalette\fun <}}
\def\ga{\mathrel{\mathpalette\fun >}}
\def\fun#1#2{\lower3.6pt\vbox{\baselineskip0pt\lineskip.9pt
        \ialign{$\mathsurround=0pt#1\hfill##\hfil$\crcr#2\crcr\sim\crcr}}}

\def\r{{\bf r}}
\def\x{{\bf x}}
\def\q{{\bf q}}
\def\k{{\bf k}}
\def\p{{\bf p}}
\def\y{{\bf y}}
\def\s{{\bf s}}
\def\u{{\bf u}}
\def\w{{\bf w}}

\def\sv{\sigma_v}
\def\su{\sigma_u}

\def\td{\tilde \delta}
\def\g2{\gamma_2}

\def\dD{\delta_{\rm D}}

\newcommand{\lexp}{\mathop{\langle}}
\newcommand{\rexp}{\mathop{\rangle}}
\newcommand{\rexpc}{\mathop{\rangle_c}}

\def\Mpc{\, h^{-1} \, {\rm Mpc}}
\def\Gpc{\, h^{-1} \, {\rm Gpc}}
\def\Gpccube{\, h^{-3} \, {\rm Gpc}^3}
\def\kvecMpc{\, h \, {\rm Mpc}^{-1}}
\def\kMpc{\, h \, {\rm Mpc}^{-1}}
\def\fnl{f_{\rm NL}}
\def\gnl{g_{\rm NL}}
\def\tnl{\tau_{\rm NL}}

\def\mnras{MNRAS}
\def\aj{AJ}
\def\apj{ApJ}
\def\apjs{ApJS}
\def\aap{A \& A}
\def\apjl{ApJ. Lett.}
\def\araa{Ann. Rev. Astron. Astrophys.}
\def\jcap{JCAP}
\def\nat{Nature}
\def\pasj{Pub. Astron. Soc. Japan}
\def\pasp{Pub. Astron. Soc. Pacific}
\def\prd{Phys. Rev. D}
\def\physrep{Phys. Rep.}
\def\qjras{QJRAS}

\title{Fast Estimators for Redshift-Space Clustering}
\author{Rom\'an Scoccimarro} \email{rs123@nyu.edu}
\affiliation{Center for Cosmology and Particle Physics, Department of Physics,  
             New York University, NY 10003, New York, USA}

\date{\today}         

\begin{abstract}
Redshift-space distortions in galaxy surveys happen along the radial direction, breaking statistical translation invariance. We construct estimators for radial distortions that, using only Fast Fourier Transforms (FFTs) of the overdensity field multipoles for a given survey geometry, compute the power spectrum monopole, quadrupole and hexadecapole, and generalize such estimators to the bispectrum. Using realistic mock catalogs we compare the signal to noise of two estimators for the power spectrum hexadecapole that require different number of FFTs and measure the bispectrum monopole, quadrupole and hexadecapole. The resulting algorithm is very efficient, e.g. for the BOSS survey requires about three minutes for $\ell=0,2,4$ power spectra for scales up to $k=0.3\kMpc$ and about fifteen additional minutes for $\ell=0,2,4$ bispectra for all scales and triangle shapes up to $k=0.2\kMpc$ on a single core. The speed of these estimators is essential as it makes possible to compute covariance matrices from large number of realizations of mock catalogs with realistic survey characteristics, and paves the way for improved constrains of gravity on cosmological scales, inflation and galaxy bias.
\end{abstract}

\maketitle

\section{Introduction}

Redshift-space distortions~\cite{Kai87,Ham92} of galaxy clustering are key in understanding the three-dimensional distribution of large-scale structure and are also a major probe for constraining gravity on cosmological scales, as evidenced in recent work~\cite{BlaGlaDav1112,de-GuzPea1309,RacBerPie1311,ReiSamWhi1211,BeuSaiSeo1409,ChuPraBeu1312,SamReiWhi1404,SanMonKaz1405}. Such distortions change the Fourier modes from their undistorted real-space values depending on the orientation of the wave-vectors with respect to the line of sight. In modern era surveys with large solid angles the line of sight is significantly space-dependent (as the radial direction varies over the sky), which makes Fourier-space analysis non-trivial beyond the lowest multipole, the monopole. 




%
%

To see this we recall that redshift-space positions $\s$ are given in terms of  real-space positions $\x$ by
\beq
\s = \x - f\, \hat{x}\, (\u\cdot\hat{x})
\label{RSDmap}
\eeq
where $f = d\ln D_+/d\ln a$ in terms of the linear growth factor $D_+$ and scale factor $a$, and the peculiar velocity $v = - {\cal H} f \,\u$ with ${\cal H} = d\ln a/ d \tau$ the comoving Hubble constant (and $\tau$ conformal time). This means that in linear perturbation theory the redshift-space density fluctuations are given by
\beq
\delta_s(\x) = \delta(\x) + f\, \nabla \cdot \Big[ \hat{x}\, (\u\cdot\hat{x}) \Big]
\label{linds}
\eeq
or $\delta_s = \delta + f \ (2\, u_r/r+\partial_ru_r)$ which when the solid angle of the survey is small enough (the so-called plane-parallel approximation, $\hat{x} \to \hat{z}$), goes to $\delta_s(\x) = \delta(\x) + f\, \nabla_z u_z$, leading to $\delta_s(\k)=(1+f \mu^2)\, \delta(\k)$ in Fourier space~\cite{Kai87}. Using that in linear theory $\delta = \nabla \cdot \u$, we can also write
\beq
\delta_s(\x) = \Big[ 1 + f \Big( \partial_r^2 + {2\over r} \partial_r \Big)\nabla^{-2}\Big]  \delta(\x) \equiv {\cal D}_s\, \delta(\x)
\label{RSDoperator}
\eeq
which gives the form of the linear redshift-space distortion operator ${\cal D}_s$. A fundamental property of radial distortions is that, unlike the plane-parallel case, the reshift-space map does not commute with translations (as the observer defines a privileged location) and thus ${\cal D}_s$ is not an eigenfunction of plane-waves, the effect of distortions on a mode is not an eigenvalue ($1+f \mu^2$) anymore, and 
as a result the Fourier power spectrum is no longer diagonal~\cite{HamCul9601}. That is, 
\beq
\langle \delta_s(\k_1) \delta_s(\k_2) \rangle = P(\k_1,\k_2)
\label{Pnonhomog}
\eeq
with $P(\k_1,\k_2)$ only becoming diagonal in the plane-parallel limit, when $P(\k_1,\k_2) \to P(\k_1)\, \dD(\k_{12})$ with $\k_{12}=\k_1+\k_2$. The fact that $P(\k_1,\k_2)$ is not diagonal is directly related to the space-dependent unit vectors in Eq.~(\ref{linds}) which makes ${\cal D}_s$ in Fourier space an integral operator inducing mode-coupling even in linear theory~\cite{ZarHof9605}. While in principle such matrix contains all the information it is not clear how to make simple use of it.  One would like to condense all the information into a set of multipoles as a function of a scalar $k$ as in the plane-parallel limit but there appears no simple way to do so. 

Of course, radial distortions preserve isotropy about the observer and thus spherical harmonics become a natural basis for angular modes. Spherical harmonics transform alternatives to Fourier analysis exist and are well known, at least for the power spectrum (e.g.~\cite{FisSchLah9401,HeaTay9507,HamCul9601,2004MNRAS.353.1201P}), although their application to surveys is not done as often due to their computational cost. For this reason in this paper we concentrate on Fourier analysis and how to tackle fast estimation of redshift-space power spectrum and bispectrum multipoles in the presence of radial distortions.

The plan for the paper is as follows. In section~\ref{localest} we now discuss a slight generalization of the power spectrum and bispectrum estimators for ``local" regions in space where the fluctuations can be taken to be approximately statistically homogeneous (and thus these local statistics can be taken as diagonal). From these in sections~\ref{Pmultipoles} and~\ref{Bmultipoles} we build multipoles estimators for the power spectrum and bispectrum respectively, as if each of these regions is in the plane-parallel approximation, giving us estimators that apply to the general case of radial distortions. In section~\ref{GalSurveys} we discuss the application to galaxy surveys and in section~\ref{conclude} we conclude.

\section{Distortions Estimators}
\label{RSDestim}

\subsection{Local Estimators}
\label{localest}

Since in the presence of radial redshift-space distortions the resulting redshift-space density field is no longer statistically homogeneous, it makes sense to define a local spectrum density estimator at $\x$,
\beq
\widehat{P}_{\rm local}(\k,\x) \equiv \int {d^3x_{12}\over (2\pi)^3}\ \delta_s(\x+\w_1)\, \delta_s(\x+\w_2)\ {\rm e}^{-i \k \cdot \x_{12}}
\label{Plocal}
\eeq
where the $\w_i$ are the coordinates of the $\x_i$ from the center of mass of the pair $\x=(\x_1+\x_2)/2$, that is $\x_i = \x + \w_i$ and $\x_{12}=\x_1-\x_2$.  Therefore $\w_1=-\x_{12}/2$, $\w_2=\x_{12}/2$. Equation~(\ref{Plocal}) is the Fourier transform of the local contribution at $\x$ to the correlation function at separation $\x_{12}$. The local power spectrum density is real but not positive definite. 

We can write Eq.~(\ref{Plocal}) in terms of Fourier coefficients,
\beq
\widehat{P}_{\rm local}(\k,\x) \equiv  \int d^3q\ \delta_s(\k+\q/2)\, \delta_s(-\k+\q/2)\ {\rm e}^{i \q \cdot \x}
\label{Ploca2l}
\eeq
Integrating over space Eq.~(\ref{Ploca2l}) over space  the local power density is simply
\beq
 \int {d^3x\over (2\pi)^3}\, \widehat{P}_{\rm local}(\k,\x) = |\delta_s(\k)|^2
\label{avgP}
\eeq
that is, the standard power spectrum estimator when it makes sense to average over space (when translation invariance holds).
Equation~(\ref{Ploca2l}) has an expectation value,
\beq
\langle \widehat{P}_{\rm local}(\k,\x) \rangle =  \int d^3q\ P(\k+\q/2,-\k+\q/2)\ {\rm e}^{i \q \cdot \x}
\label{expPlocal}
\eeq
where we have used that the power spectrum in the presence of radial distortions is no longer diagonal due to the loss of statistical homogeneity or translation invariance. In the case that there is statistical translation invariance (e.g. in the plane-parallel limit), $P(\k+\q/2,-\k+\q/2) = P(\k)\, \dD(\q)$ and we have that $\langle \widehat{P}_{\rm local}(\k,\x) \rangle = P(\k)$, as expected.  But note that in general, $\langle \widehat{P}_{\rm local}(\k,\x)\rangle $ contains the same information as the matrix $P(\k_1,\k_2)$, as the latter can be recovered from the former by a Fourier transform. The advantage of the former now becomes obvious, as there is a trace of real space (and thus line of sight) that can be used to take multipoles and thus compress the information on redshift distortions in a similar way as in the plane-parallel limit, as we discuss in the next section. 

We can now extend these definitions to write down a local bispectrum estimator ($\k_3=-\k_{12}$)
\beqa
\widehat{B}_{123}^{\rm local}(\x) &\equiv& \int {d^3x_{13}\over (2\pi)^3}\, {d^3x_{23}\over (2\pi)^3} 
\ {\rm e}^{-i (\k_1 \cdot \x_{13}+ \k_2 \cdot \x_{23})} 
\nonumber \\ & & \times \prod_{i=1}^3\, \delta_s(\x+\w_i)
\label{Blocal}
\eeqa
where the $\w_i$ are the vectors from the centroid of the triangle $\x$ to the vertices $\x_i$, i.e. $\x=(\x_1+\x_2+\x_3)/3$ with $\x_i = \x + \w_i$ and $\x_{ij}\equiv \x_i-\x_j$.  The centroid coordinates of the vertices are $\w_1=2\, \x_{13}/3-\x_{23}/3$, $\w_2=2\, \x_{23}/3-\x_{13}/3$, and $\w_3=-\x_{13}/3-\x_{23}/3$. In terms of Fourier coefficients,
\beqa
\widehat{B}_{123}^{\rm local}(\x) &=& \int d^3q \ {\rm e}^{-i \q\cdot\x}\, \prod_{i=1}^3\, \delta_s(\k_i+\q/3)
\label{Blocal2}
\eeqa
with expectation value
\beq
\langle \widehat{B}_{123}^{\rm local}(\x) \rangle = \int d^3q \ B(\k_1+\q/3,\k_2+\q/3,\k_3+\q/3)\, {\rm e}^{-i \q\cdot\x} 
\label{BlocalEV}
\eeq
which involves the (non-diagonal) bispectrum (note the similarity with Eq.~\ref{expPlocal}). Again, as in the power spectrum case, when it does make sense to spatially average, we have
\beq
\int {d^3x \over (2\pi)^3} \ \widehat{B}_{123}^{\rm local}(\x) = \prod_{i=1}^3\, \delta_s(\k_i)
\label{BlocalAvg}
\eeq
the standard bispectrum estimator for a statistically homogeneous field.

\subsection{Power Spectrum Multipoles}
\label{Pmultipoles}

From Eq.~(\ref{Ploca2l}) we can build the natural estimator of power multipoles by using $\hat{x}$ as the line of sight direction to the pair of points,
\beq
\widehat{P}_\ell(k) \equiv (2\ell+1) \int {d\Omega_k \over 4\pi}  \int {d^3 x \over (2\pi)^3}\ \widehat{P}_{\rm local}(\k,\x)\, {\cal L}_\ell(\hat{k}\cdot\hat{x})
\label{Pell_local}
\eeq
where by angular integration we actually mean integration over a thin shell in Fourier space centered at $k$, for any function $F$
\beq
\int {d\Omega_k \over 4\pi}\, F(\k) \equiv \int_k {d^3q \over N_k}\, F(\q)
\label{shell}
\eeq
where $N_k\equiv \int_k d^3q = 4\pi k^2 \delta k$ is the volume of the shell in $k$-space, with $\delta k$ the bin size. Equation~(\ref{Pell_local}) can be rewritten using Eq.~(\ref{Plocal}) as,
\beqa
\widehat{P}_\ell(k) &=& {(2\ell+1)} \int {d\Omega_k \over 4\pi} \int {d^3 x_1 \over (2\pi)^3} {d^3 x_2 \over (2\pi)^3} 
 {\rm e}^{-i \k \cdot \y}
\, {\cal L}_\ell(\hat{k}\cdot\hat{x}) \nonumber \\  & & \times\  \delta_s(\x_1)\delta_s(\x_2)\, 
\label{Pell_local2}
\eeqa
where $\y = \x_1-\x_2$ and $\x=\x_1+\x_2$. This natural estimator is for $\ell=2$ precisely the same as the so-called Yamamoto estimator~\cite{YamNakKam0602}. The complication with such estimators is well-known, i.e. they are expensive to compute beyond $\ell=0$ because the integrals in Eq.~(\ref{Pell_local2}) do not decouple into a product of Fourier transforms due to the $\x$ dependence in ${\cal L}_\ell(\hat{k}\cdot\hat{x})$, and thus when discretized the integrals become double sums  that are quadratic in the number of grid points (or galaxies), leading to an $N^2$ bottleneck. 

Given this complication, it has been proposed~\cite{BlaDavPoo1108,BeuSaiSeo1409} to make the replacement for the line of sight definition,
\beq
{\cal L}_\ell(\hat{k}\cdot\hat{x}) \to {\cal L}_\ell(\hat{k}\cdot\hat{x_1})
\label{1LOS}
\eeq
to make the two integrals (or sums in the discrete case) factorize. Still, without further treatment, one the integrals (or sums) which contains the Legendre polynomial is not of Fourier form due to the $\hat{k}$ dependence for $\ell>0$. While faster than the natural estimator which requires dealing with pairs of points, this is still expensive to compute compared to the power spectrum monopole which can just be computed with a single Fourier transform. The replacement in Eq.~(\ref{1LOS}) has recently been found to be rather accurate compared to the natural estimator by~\cite{SamBraPer1504}, and the latter accurate compared to the full description of the power spectrum including wide-angle effects~\cite{YooSel1308}, so it is of importance to find a fast way to compute it.

One of the main points of this paper is to point out that while the replacement in Eq.~(\ref{1LOS}) cannot be written as a product of Fourier transforms, it can in fact be written as the sum of a product of Fourier transforms, and as a result of this,  estimators can be built that are computable using a handful of FFTs.  Indeed, as a result of this replacement, Eq.~(\ref{Pell_local2}) becomes the cross correlation of the local multipole overdensity with the local monopole~\cite{BlaDavPoo1108,BeuSaiSeo1409,SamBraPer1504} where
\beq
\delta_\ell(\k) \equiv \int {d^3 x \over (2\pi)^3} {\rm e}^{-i\k\cdot\x}\, \delta_s(\x)\, {\cal L}_\ell(\hat{k}\cdot\hat{x})
\label{deltaell}
\eeq
which in fact {\em can} be computed by FFTs by simply factorizing out the $\hat{k}$-dependence, e.g. for $\ell=2$
\beq
\delta_2(\k) = {3\over 2} \hat{k}_i \hat{k}_j Q_{ij}(\k) -{1\over 2} \delta_0(\k)
\label{delta2}
\eeq
with
\beq
Q_{ij}(\k) \equiv \int {d^3 x \over (2\pi)^3} {\rm e}^{-i\k\cdot\x}\, \delta_s(\x)\, \hat{x}_i \hat{x}_j
\label{Qij}
\eeq
which depends on volume shape through the cosines $\hat{x}_i$ (as the integral in Eq.~\ref{Qij} for a survey becomes over the region where $\delta_s$ is observed, 
see section~\ref{GalSurveys} below). Since this is a symmetric tensor only 6 FFTs are needed to compute it in the absence of any symmetry of the survey geometry, i.e.
\beqa
\delta_2 &=& {3\over 2} \Big[ 
\hat{k}_x^2 Q_{xx} +\hat{k}_y^2 Q_{yy} +\hat{k}_z^2 Q_{zz}  \nonumber \\ & & 
+2\hat{k}_x \hat{k}_y Q_{xy} +2\hat{k}_y \hat{k}_z Q_{yz}+2\hat{k}_z \hat{k}_x Q_{zx}
\Big] 
-{1\over 2} \delta_0 \nonumber \\ &&
\label{delta2b}
\eeqa
In the plane-parallel limit, all $Q_{ij}$ vanish but $Q_{zz}$ and we recover the standard plane-parallel results. 
For $\ell=4$ we have, similarly 
\beq
\delta_4(\k) = {35\over 8} \hat{k}_i \hat{k}_j \hat{k}_l \hat{k}_k Q_{ijlk}(\k) -{5\over 2} \delta_2(\k)-{7\over 8} \delta_0(\k)
\label{delta4}
\eeq
with
\beq
Q_{ijlk}(\k) \equiv \int {d^3 x \over (2\pi)^3} {\rm e}^{-i\k\cdot\x}\, \delta_s(\x)\, \hat{x}_i \hat{x}_j\hat{x}_l \hat{x}_k
\label{Qijkl}
\eeq
and so on. Since $Q_{ijlk}$ is a fully symmetric tensor, in the absence of any symmetries, one needs to compute 15 FFTs to fully characterize it,
\beqa
\delta_4 &=& {35\over 8} \Big[ 
\hat{k}_x^4\, Q_{xxxx} + {\rm (3)~cyc.}
+4\hat{k}_x^3 \hat{k}_y\, Q_{xxxy} +{\rm (6)~cyc.}  \nonumber \\ & & 
+ 6 \hat{k}_x^2 \hat{k}_y^2\, Q_{xxyy} +{\rm (3)~cyc.}
+ 12 \hat{k}_x^2 \hat{k}_y \hat{k}_z\, Q_{xxyz}  \nonumber \\ & & +{\rm (3)~cyc.}
\Big] 
-{5\over 2} \delta_2 - {7\over 8} \delta_0 \nonumber \\ &&
\label{delta4b}
\eeqa
where the number in parenthesis denotes how many total terms belong to each cyclic permutation (and counts the numbers of FFTs needed).  Again, in the plane-parallel limit, all $Q_{ijkl}$ vanish but $Q_{zzzz}$ and we recover the standard plane-parallel results.

The power spectrum multipoles estimator then would be~\cite{BlaDavPoo1108,BeuSaiSeo1409,SamBraPer1504}

\beq
\widehat{P}_\ell(k) = {(2\ell+1)} \int {d\Omega_k \over 4\pi}\, \delta_\ell(\k)\, \delta_0(-\k)
\label{Pell_local3}
\eeq
and thus for $\ell=0,2,4$ a total of $1+6+15=22$ FFTs would be needed. Because of $N \log N$ scaling even this many FFTs is still many orders of magnitude faster than the naive $N^2$ procedure where the Legendre polynomials are not written in factorized form. 

One obvious question that arises is whether one can do better for the hexadecapole (and higher multipoles) as it is rather disappointing that one needs so much additional computational cost (an additional 15 FFTs for $\ell=4$) to describe a quantity that has rather poor signal to noise compared to $\ell=0,2$ at large scales. The answer is that one {\em can do better} by not constraining oneself to a single line of sight as $\ell>2$ is considered. To see this let us consider the case $\ell=4$ for definiteness. We go back to the natural estimator in Eq.~(\ref{Pell_local2}) and write the fourth-order Legendre polynomial in terms of quadratic combinations of lower even multipoles,

\beq
{\cal L}_4(\hat{k}\cdot \hat{x}) = {35 \over 18} \Big[{\cal L}_2(\hat{k}\cdot \hat{x})\Big]^2-{5 \over 9} {\cal L}_2(\hat{k}\cdot \hat{x})
-{7 \over 18} 
\label{factorizeL4}
\eeq
and then we simply split the first term 
\beq
\Big[{\cal L}_2(\hat{k}\cdot \hat{x})\Big]^2 \to {\cal L}_2(\hat{k}\cdot \hat{x}_1) \ {\cal L}_2(\hat{k}\cdot \hat{x}_2)
\label{split}
\eeq
leading to the factorized hexadecapole estimator,
\beq
\widehat{P}_{4b}(k) =  {35\over 2} \int {d\Omega_k\over 4\pi} 
 |\delta_2(\k)|^2 -  \widehat{P}_2(k) - {7\over 2}\, \widehat{P}_0(k)
\label{P4fast}
\eeq
which does not require any additional FFTs over those already computed for $\ell=0,2$ and is based on the autocorrelation of the local quadrupole of overdensities generated by the redshift-space mapping. As a result of this split, computing $\ell=0,2,4$ requires only 7 FFTs instead of 22, leading to additional computational savings of just over a factor of three by using $\widehat{P}_{4b}$ instead of $\widehat{P}_{4}$. In~\cite{SamBraPer1504} it was found that the estimator $\widehat{P}_{4}$ in Eq.~(\ref{Pell_local3}) has a small bias compared to the natural estimator at large scales, we study the bias of $\widehat{P}_{4b}$ relative to $\widehat{P}_{4}$ and their cosmic variance in section~\ref{GalSurveys} below, and find that $\widehat{P}_{4}$ is preferred due to lower cosmic variance, although the difference might be negligible for future surveys. 

The same line-of-sight split trick can be used for higher multipoles, in the obvious way. Note that for $\ell=6$ one cannot avoid computing the 15 FFTs, but this also allows one to compute $\ell=8$ without extra FFTs. In general each new set of FFTs becomes useful for two multipoles, as the Legendre polynomials can be split in quadratic combinations because of the two line of sights available in a two-point function. 

\subsection{Bispectrum Multipoles}
\label{Bmultipoles}

We now consider the case of the bispectrum. In the plane-parallel limit, the bispectrum becomes a function of five variables: the three sides plus two angular variables describing the orientation of the triangle with respect to the line of sight. A third angular variable is irrelevant in the sense that it rotates the triangle about the line of sight, leaving the redshift-space bispectrum invariant. A convenient way to handle the plane-parallel case is then to do a spherical harmonic decomposition with respect to the two relevant angular variables.  Let $k_1 \geq k_2 \geq k_3$ without loss of generality. From the local bispectrum estimator in Eq.~(\ref{Blocal}) we can then define the multipoles in the radial distortions case by
\beqa
\widehat{B}_{123}^{(\ell m)} &\equiv& {(2\ell+1)\over N^{\rm T}_{123}} \prod_{i=1}^3 \int_{k_i} d^3\q_i  \, \dD(\q_{123}) \nonumber \\
& & \times \int {d^3x \over (2\pi)^3} \, 
\widehat{B}_{123}^{\rm local}(\x)\ Y_{\ell m}(\theta_1,\phi_{12})
\label{BispLM}
\eeqa
where the indices 1,2,3 in $\widehat{B}_{123}^{\rm local}$ now refer to the $\q_i$, which are being averaged over a shell of thickness $\delta k$ about the $k_i$, i.e. $\int_{k_i} d^3 q_i= \int N_{k_i} (d\Omega_{k_i}/4\pi)$ for thin shells, 
and 
\beq
N^{\rm T}_{123} =  \prod_{i=1}^3 \, \int_{k_i} {d^3q_i}\, \dD(\q_{123}) 
\simeq  8\pi^2\, k_1k_2k_3 \, \delta k^3
\label{NT}
\eeq
In Eq.~(\ref{BispLM}) we followed~\cite{ScoCouFri9906} where $\cos\theta_i \equiv \hat{q}_i\cdot\hat{x}$, $\cos \theta_{12} \equiv \hat{q}_1\cdot \hat{q}_2$ and $\phi_{12}$ is the azimuthal angle of $\q_2$ around $\q_1$ satisfying $\cos\theta_2=  \cos\theta_1 \cos \theta_{12} - \sin\theta_1 \sin \theta_{12} \cos \phi_{12}$.  Another possible choice of angular variables are those that describe the orientation of the normal to the triangle face. However, it has the disadvantage that it is not well defined for zero area triangles but this can be handled separately as for such triangles a Legendre multipole decomposition is all that is needed as all $\hat{q}_i$ differ at most by a sign (irrelevant for even multipoles). 

Here for simplicity we take Legendre multipoles with respect to the largest side, corresponding to $m=0$ multipoles in Eq.~(\ref{BispLM}), that is 
\beqa
\widehat{B}_{123}^{(\ell)} &\equiv& {(2\ell+1)\over N^{\rm T}_{123}} \prod_{i=1}^3 \int_{k_i} d^3\q_i  \, \dD(\q_{123}) \nonumber \\
& & \times \int {d^3x \over (2\pi)^3} \, 
\widehat{B}_{123}^{\rm local}(\x)\ {\cal L}_\ell(\hat{q}_1\cdot\hat{x})
\label{YBisp}
\eeqa
which can also be written as,
\beqa
\widehat{B}_{123}^{(\ell)} &=& {(2\ell+1)\over N^{\rm T}_{123}} \, \prod_{i=1}^3\, \int_{k_i} d^3\q_i  \, \dD(\q_{123}) \nonumber \\
& & \times \int {d^3x_i \over (2\pi)^3}\ 
\delta(\x_i) \, {\rm e}^{-i \q_i \cdot \x_i} \, {\cal L}_\ell(\hat{q}_1\cdot\hat{x})
\label{YBisp2}
\eeqa
which is obviously in the same ``natural" form as for the power spectrum. We must now deal with the separability of the estimator, as in the power spectrum case. For $\ell=2$ using $\hat{x} \to \hat{x}_1$ we obtain the factorized estimator for  the bispectrum quadrupole,
\beq
\widehat{B}^{(2)}_{123} \equiv  5\,  \prod_{i=1}^3\, \int_{k_i} d^3\q_i  \, {\dD(\q_{123}) \over N^{\rm T}_{123}}\ 
 \delta_2(\q_1)\, \delta_0(\q_2)\, \delta_0(\q_3)
\label{Bquad}
\eeq
and so computing the bispectrum quadrupole only results in a total factor of two over monopole alone, as the computation of $\delta_2$ is negligible in cost with the bispectrum itself (and is the same ingredient needed for the power spectrum quadrupole). A similar consideration leads to the hexadecapole bispectrum estimator,
\beq
\widehat{B}^{(4)}_{123} \equiv  9\,  \prod_{i=1}^3\, \int_{k_i} d^3\q_i  \, {\dD(\q_{123}) \over N^{\rm T}_{123}}\ 
 \delta_4(\q_1)\, \delta_0(\q_2)\, \delta_0(\q_3)
\label{Bispell4}
\eeq
and the obvious generalization for higher-order multipoles. As discussed above for the power spectrum, additional multipole information may be obtained instead by including more than one quadrupole field $\delta_2$ in Eq.~(\ref{Bquad}) instead of computing the additional 15 FFTs to build $\delta_4$. That is, if for $\ell=4$ we may use Eq.~(\ref{YBisp2}) and Eq.~(\ref{factorizeL4}) with the split
\beq
\Big[{\cal L}_2(\hat{q}_1\cdot \hat{x})\Big]^2 \to {\cal L}_2(\hat{q}_1\cdot \hat{x}_1) \ {\cal L}_2(\hat{q}_1\cdot \hat{x}_2)
\label{splitB}
\eeq
we obtain the alternative hexadecapole bispectrum estimator,
\beqa
\widehat{B}^{(4b)}_{123} &\equiv &{35\over 2} \prod_{i=1}^3\, \int_{k_i} d^3\q_i  \, {\dD(\q_{123}) \over N^{\rm T}_{123}}\ 
 \delta_2(\q_1)\, \delta_2(\hat{q}_1,\q_2)\, \delta_0(\q_3) \nonumber \\ & & 
-  \widehat{B}^{(2)}_{123} - {7\over 2}\, \widehat{B}^{(0)}_{123}
\label{Bhexa}
\eeqa
where
\beq
\delta_2(\hat{p},\q) \equiv {3\over 2}\, \hat{p}_i\hat{p}_j \, Q_{ij}(\q) -{1\over 2} \delta_0(\q)
\label{mixedMultipole}
\eeq
Note that for zero area triangles $\delta_2(\hat{q}_1,\q_2)=\delta_2(\q_2)$ and thus,
\beqa
\widehat{B}^{(4)}_{123} &\equiv &{35\over 2} \prod_{i=1}^3\, \int_{k_i} d^3\q_i  \, {\dD(\q_{123}) \over N^{\rm T}_{123}}\ 
 \delta_2(\q_1)\, \delta_2(\q_2)\, \delta_0(\q_3) \nonumber \\ & & 
-  \widehat{B}^{(2)}_{123} - {7\over 2}\, \widehat{B}^{(0)}_{123}
\label{Bhexa2}
\eeqa
which is  analogous to the power spectrum case Eq.~(\ref{P4fast}). A disadvantage of the estimator in Eq.~(\ref{Bhexa}) is that the extra $\hat{q}_1$ dependence means that the bispectrum estimator is a bit more costly, as e.g. for $\ell=4$ we must now estimate
\beq
\prod_{i=1}^3\, \int_{k_i} d^3\q_i  \, {\dD(\q_{123})}\ 
(\hat{q}_1)_i (\hat{q}_1)_j\, \delta_2(\q_1)\, Q_{ij}(\q_2)\, \delta_0(\q_3) 
\label{ell4complication}
\eeq
which corresponds to evaluating six bispectra.  Since the cost of evaluating the bispectrum is much more than evaluating the next set of 15 FFTs of the overdensity fields in Eq.~(\ref{delta4b}), it is more convenient in this case to avoid the split in Eq.~(\ref{splitB}) and instead use Eq.~(\ref{Bispell4}).

Another approach would be to use the zero-area triangle estimator in Eq.~(\ref{Bhexa2}) for {\em all} triangles. Unlike Eq.~(\ref{Bispell4}), this does not need the extra 15 FFTs and it is as fast to estimate. However, this estimator for general triangles is not a true multipole, that is, it does not vanish in real space except for zero area triangles and therefore extracting information from it about redshift-space distortions may be more complicated due to degeneracies with the monopole. However, it may be useful  not just for zero area triangles, but at bit more generally for nearly squeezed or folded triangles. It is beyond the scope of this paper to explore this further, in what follows we will consider the more standard bispectrum multipoles as in Eq.~(\ref{Bquad}) and~(\ref{Bispell4}), particularly in light of the reduced cosmic variance of such estimators compared to those that require less FFTs as we find below for the power spectrum multipoles.

Finally, for completeness we briefly mention how to build the zero-area triangle estimator for $\ell=6$. We split the sixth-order Legendre polynomial in cubic combinations of lower-order even polynomials, as we now have three lines of sight. Using that 
\beq
{\cal L}_6(\hat{k}\cdot\hat{x}) = 
{77\over 18} \Big[{\cal L}_2(\hat{k}\cdot\hat{x})\Big]^3 - {7\over 3} \Big[{\cal L}_2(\hat{k}\cdot\hat{x})\Big]^2 - 
 {7\over 6} {\cal L}_2(\hat{k}\cdot\hat{x})+ {2\over 9}
 \label{factorizeL6}
 \eeq
we simply split the first term 
\beq
\Big[{\cal L}_2(\hat{q}_1\cdot \hat{x})\Big]^3 \to {\cal L}_2(\hat{q}_1\cdot \hat{x}_1) \ {\cal L}_2(\hat{q}_2\cdot \hat{x}_2)\ {\cal L}_2(\hat{q}_3\cdot \hat{x}_3)
\label{splitCube}
\eeq
which leads to,
\beqa
\widehat{B}^{(6)}_{123} &\equiv &{1001\over 18} \prod_{i=1}^3\, \int_{k_i} d^3\q_i  \, {\dD(\q_{123}) \over N^{\rm T}_{123}}\ 
 \delta_2(\q_1)\, \delta_2(\q_2) \, \delta_2(\q_3) \nonumber \\ & &  \times  
+ {26 \over 15}  \widehat{B}^{(4)}_{123}- {13\over 10} \widehat{B}^{(2)}_{123} + {283\over 45}\, \widehat{B}^{(0)}_{123}
\nonumber \\ & & 
\label{Bell6}
\eeqa
so now with the six FFTs needed for the quadrupole we can compute up to the $\ell=6$ zero-area triangle bispectrum multipole,  so each set of FFTs is enough for three multipoles as there are three lines of sight in a three-point function.

\section{Implementation in Galaxy Surveys}
\label{GalSurveys}

\subsection{Power Spectrum}

We now proceed to implementing the above ideas in the case of a survey geometry, where the galaxy sample is characterized by $N_g$ galaxies at positions $\x_j$ and the radial and angular selection functions by a random catalog with $N_r$ objects ($\alpha \equiv N_g/N_r \ll 1$). Each object is given a weight $w_j$, e.g. the FKP weights~\cite{FelKaiPea9405}. Given the results above, we  write the overdensity monopole
\beq
F_0(\k) \equiv \Big( \sum_{j=1}^{N_g} - \alpha \sum_{j=1}^{N_r} \Big) w_j \, {\rm e}^{i \k \cdot \x_j}
\label{F0}
\eeq
whereas for the quadrupole we have
\beq
F_2(\k) \equiv {3\over 2} \hat{k}_a \hat{k}_b Q^{ab}(\k) -{1\over 2} F_0(\k)
\label{F2}
\eeq
with
\beq
Q^{ab}(\k) \equiv \Big( \sum_{j=1}^{N_g} - \alpha \sum_{j=1}^{N_r} \Big) \hat{x}_j^a \hat{x}_j^b\ w_j \, {\rm e}^{i \k \cdot \x_j}
\label{Qab}
\eeq


Thus the power spectrum multipoles estimators for $\ell=0,2,4$ are
\beq
\widehat{P}_0(k) = {1\over I_{22}} \Big[  \int {d\Omega_k \over 4\pi}\ |F_0(\k)|^2 - N_0  \Big]
\label{P0_impl}
\eeq

\beq
\widehat{P}_2(k) = {5\over I_{22}}  \int {d\Omega_k \over 4\pi}\ F_2(\k) F_0^*(\k)  
\label{P2_impl}
\eeq

\beq
\widehat{P}_4(k) = {9\over  I_{22}}  \int {d\Omega_k \over 4\pi}\ F_4(\k) F_0^*(\k)
\label{P4_impl}
\eeq

\beq
\widehat{P}_{4b}(k) = {7\over  10\, I_{22}}  \int {d\Omega_k \over 4\pi}\ |F_2(\k)|^2 -  \widehat{P}_2(k) - {7\over 2}\, \widehat{P}_0(k)
\label{P4_implb}
\eeq

where the standard normalization constant is~\cite{FelKaiPea9405}
\beq
I_{22} \equiv \alpha \sum_{j=1}^{N_r} \bar{n}(\x_j)\, w_j^2
\label{I22def}
\eeq
and the shot noise obtained from the self-pairs in the first term in Eq.~(\ref{P0_impl})
\beq
N_0 \equiv  \Big( \sum_{j=1}^{N_g} + \alpha^2 \sum_{j=1}^{N_r} \Big) \, w_j^2
\label{N0true}
\eeq
with the first term representing the true shot noise of galaxies, the second that of random objects. Replacing the true galaxy shot noise by its expectation value $ \sum_{j=1}^{N_g} \to  \alpha \sum_{j=1}^{N_r}$ leads to $N_0 = \alpha (1+\alpha) \sum_{j=1}^{N_r} \, w_j^2$, the standard result~\cite{FelKaiPea9405}. However, using the true shot noise should  always be preferred as it uses more information about the data~\cite{Ham00}. Note that in general one would have a  shot noise 
\beq
N_\ell \equiv  \Big( \sum_{j=1}^{N_g} + \alpha^2 \sum_{j=1}^{N_r} \Big) \, w_j^2 \int {d\Omega_k \over 4\pi} {\cal L}_\ell(\hat{k}\cdot \hat{x}_j)
\label{Nelltrue}
\eeq
but this vanishes for $\ell>0$. 

The implementation of the above sums by FFTs is straightforward and follows standard practice, e.g. the objects (galaxies or random) are interpolated to a grid to obtain a number density estimator $n$ at gridpoints $\x$, so that for any (tensor) quantity $T$ 
\beq
\Big( \sum_{j=1}^{N_g} - \alpha \sum_{j=1}^{N_r} \Big) T_j\, {\rm e}^{i \k \cdot \x_j} \to 
\sum_{\x} \Big[ n_g(\x)-\bar{n}(\x)  \Big] T(\x)\, {\rm e}^{i \k \cdot \x}
\label{mapFFT}
\eeq
where $\bar{n} = \alpha\, n_r$ and the sum over the gridpoints $\x$ is performed using FFTs. In our implementation we use fourth-order interpolation to interlaced grids, which  has superb  antialiasing properties~\cite{HocEas89}. Compared to simpler interpolations, e.g. second-order (cloud in cell), fourth-order interpolation costs a factor of 8 more in computational time and interlacing another factor of 2, but this allows us to go up to the Nyquist frequency without any significant bias (thus saving a factor of at least eight due to the reduced size of the FFT to reach the same physical $k$). A detailed analysis of interpolation techniques and their impact on clustering properties will be presented elsewhere~\cite{SefCroSco1509}. 

\begin{figure}
 \centering
 \includegraphics[ width=0.95 \linewidth]{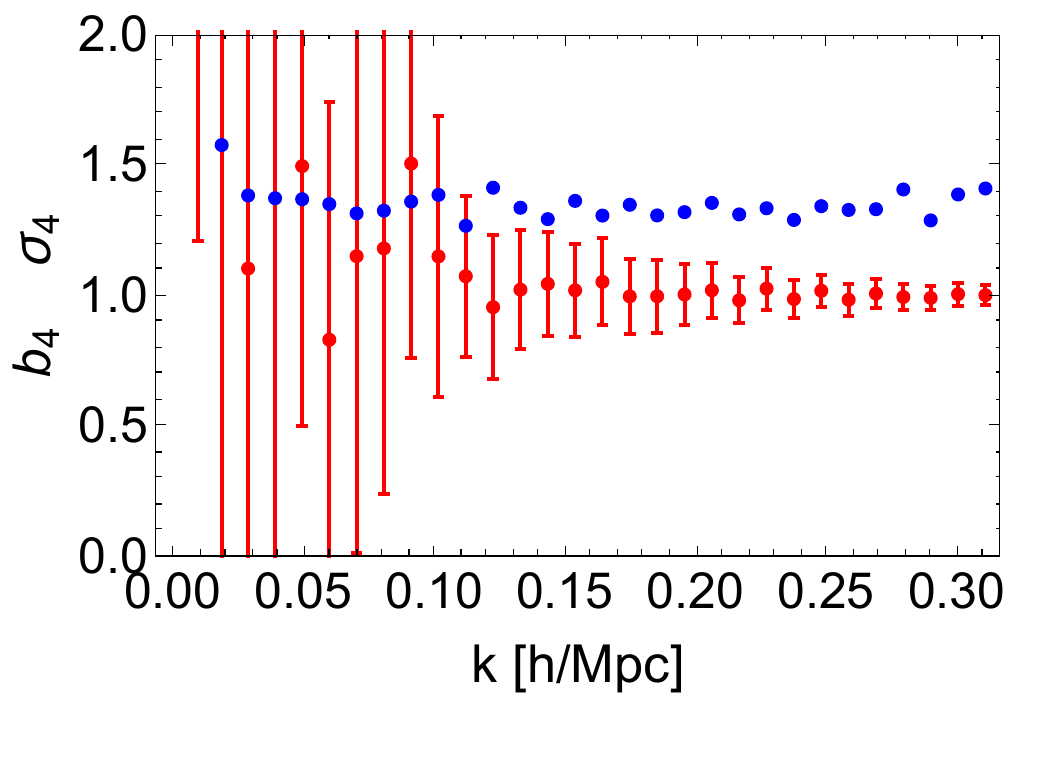}
 \caption{The bias ratio $b_4 \equiv \langle \widehat{P}_{4b}\rangle/\langle\widehat{P}_{4}\rangle$ between the two estimators of the hexadecapole (symbols with error bars), and the ratio of their cosmic variance $\sigma_4^2\equiv \langle \Delta\widehat{P}_{4b}^2\rangle/\langle\Delta\widehat{P}_{4}^2\rangle$ as a function of $k$ for the Las Damas LRG ($M_g < -21.8$) DR7 mock catalogs.}
 \label{P4LD}
\end{figure}

\begin{figure}
 \centering
 \includegraphics[ width=0.95 \linewidth]{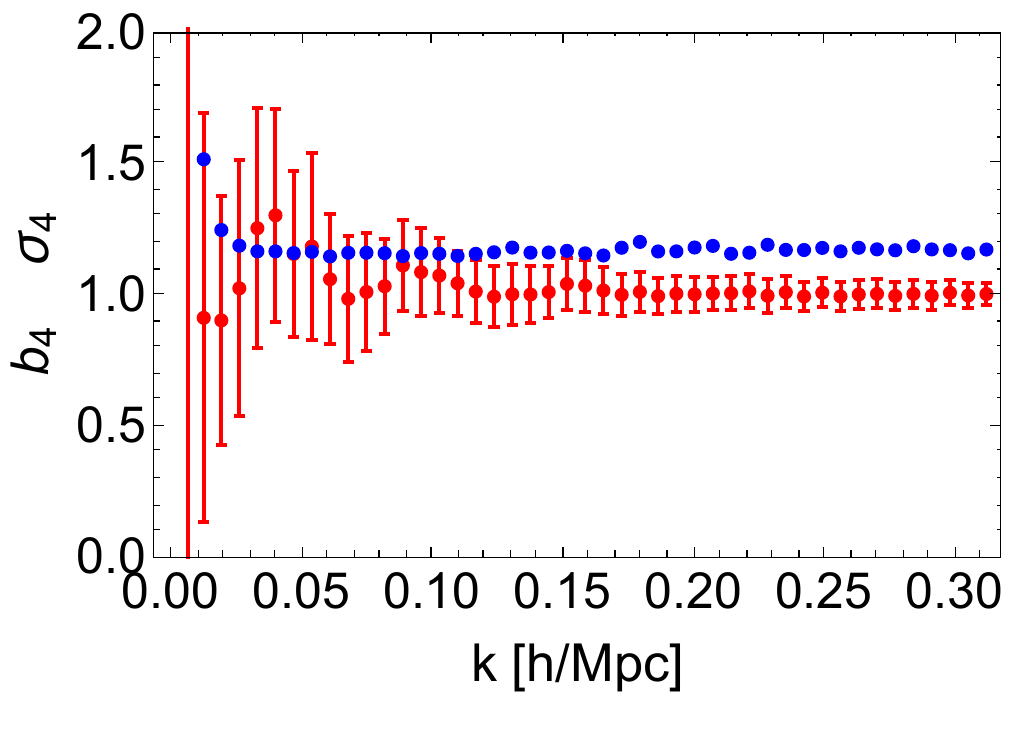}
 \caption{Sames as Fig.~\protect\ref{P4LD} for the PTHalos CMASS DR11 mock catalogs.}
 \label{P4PTH}
\end{figure}

We now compare the two hexadecapole estimators $\widehat{P}_{4}$ and $\widehat{P}_{4b}$ in terms of their expectation value and cosmic variance. For this purpose we use two sets of mock catalogs: the Las Damas LRG ($M_g < -21.8$) DR7 mocks catalogs\footnote{publicly available at {\sf http://lss.phy.vanderbilt.edu/lasdamas}.} (160 realizations) and the PTHalos CMASS DR11 mocks catalogs\footnote{publicly available at {\sf http://www.marcmanera.net/mocks}. We use version 11.1, see \protect\cite{ManScoPer1301} for more details. } (600 realizations). In both cases we only use their north galactic cup versions. Figures~\ref{P4LD} and~\ref{P4PTH} shows the results. We see that while both estimators agree within the errors on their expectation value, the cosmic variance of $\widehat{P}_{4}$  is smaller than for $\widehat{P}_{4b}$ with the difference being smaller for the higher redshift and more dense sample (CMASS). Therefore the more computationally expensive $\widehat{P}_{4}$ is preferred over the cheaper $\widehat{P}_{4b}$. This is not a very significant shortcoming as $\widehat{P}_{4}$ is still orders of magnitude faster than traditional $N^2$ estimates.

\subsection{Bispectrum}

The bispectrum multipoles estimator is given by, 
\beq
\widehat{B}^{(0)}_{123}  = \prod_{i=1}^3 \, \int_{k_i} {d^3q_i}\, {\dD(\q_{123}) \over N^{\rm T}_{123}\ I_{33}  }
 F_0(\q_1) F_0(\q_2) F_0(\q_3) - N^{(0)}_{123}
\label{BispMono}
\eeq
where  the shot noise term is given by,
%
\beqa
 N^{(0)}_{123} &=&
\prod_{i=1}^3 \, \int_{k_i} {d^3q_i}\, {\dD(\q_{123}) \over N^{\rm T}_{123}\ I_{33}  }
\Big[ F_0(\q_1) F^w_0(-\q_1)  + {\rm cyc.} \Big] 
\nonumber \\ & & 
- {2\over I_{33}} \Big( \sum_{j=1}^{N_g} - \alpha^3 \sum_{j=1}^{N_r} \Big) \, w_j^3
\label{Noise0}
\eeqa
and for higher multipoles we obtain,
\beqa
\widehat{B}^{(\ell)}_{123}  & = & (2\ell+1) \prod_{i=1}^3 \, \int_{k_i} {d^3q_i}\, {\dD(\q_{123}) \over N^{\rm T}_{123}\ I_{33}  }
 F_\ell(\q_1) F_0(\q_2) F_0(\q_3) \nonumber \\ & & - N^{(\ell)}_{123}
\label{BispQuad}
\eeqa
with the shot noise given by ($\ell >0$),
\beqa
 N^{(\ell)}_{123} &=&
(2\ell+1) \prod_{i=1}^3 \, \int_{k_i} {d^3q_i}\, {\dD(\q_{123}) \over N^{\rm T}_{123}\ I_{33}  } 
\Big[ F_\ell(\q_1) F^w_0(-\q_1) \nonumber \\ & & 
+ F_0(\q_2) F^w_\ell(\hat{q}_1,-\q_2) + F_0(\q_3) F^w_\ell(\hat{q}_1,-\q_3) \Big] \nonumber \\ & &
 \label{SNbisp2}
\eeqa
where 
\beq
F_\ell^w(\hat{q}_1,\q) \equiv
\Big( \sum_{j=1}^{N_g} + \alpha^2 \sum_{j=1}^{N_r} \Big) \, w_j^2\, 
{\rm e}^{i \q\cdot \x_j}\, {\cal L}_\ell(\hat{q}_1\cdot \hat{x}_j)
\eeq
and $F_\ell^w(\hat{q}_1,\q_1)\equiv F_\ell^w(\q_1)$. If desired the estimator in Eq.~(\ref{BispQuad}) can be symmetrized over its arguments in the obvious way. In the plane-parallel limit, this estimator for $\ell=2$ reduces to the one in~\cite{ScoCouFri9906}, which was used to measure the bispectrum quadrupole in Nbody simulations.

\begin{figure}
 \centering
 \includegraphics[ width=0.95 \linewidth]{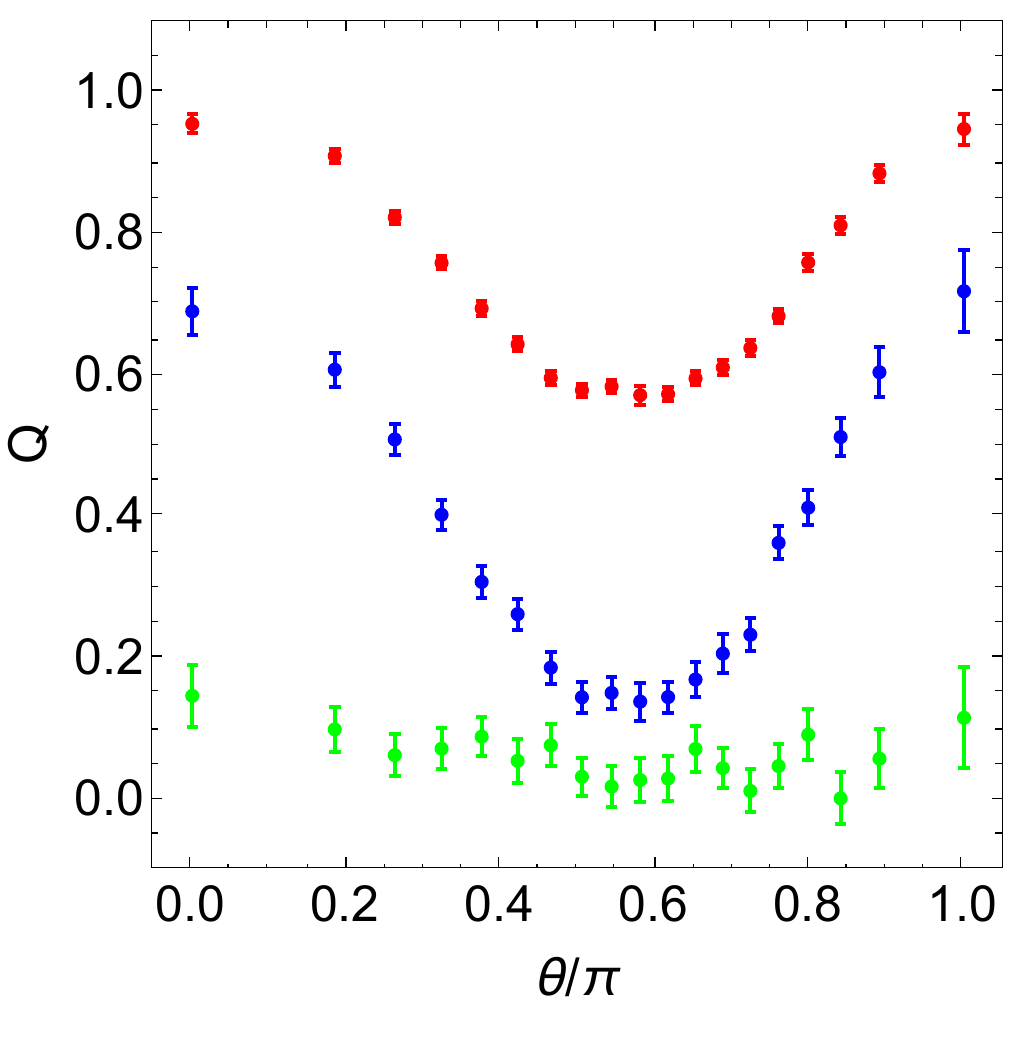}
 \caption{Reduced bispectrum multipoles $Q_{123}^{(\ell)}$ ($\ell=0,2,4$ from top to bottom) for galaxies in the LasDamas LRG ($M_g<-21.8$) DR7 mock catalogs. The triangles correspond to $k_1=0.047 \kMpc$, $k_2=2\, k_1$ as a function of the angle $\theta$ between $\k_1$ and $\k_2$.}
 \label{Qell}
\end{figure}

We can simplify Eq.~(\ref{SNbisp2}) by assuming the thin-shell approximation, which leads to
\beqa
 N^{(\ell)}_{123}  &\simeq & 
 (2\ell+1) \int_{k_1} {d^3q_1\over N_{k_1}}\, { F_\ell(\q_1) F^w_0(-\q_1) \over I_{33}  } + \nonumber \\ & &  
 (2\ell+1) \sum_{i=2}^3 {\cal L}_\ell(\hat{q}_1 \cdot \hat{q}_i) \int_{k_i} {d^3q_i\over N_{k_i}}\, { F_0(\q_i) F^w_\ell(-\q_i) \over I_{33}  } \nonumber \\ & &
 - {2\over I_{33}} \Big( \sum_{j=1}^{N_g} - \alpha^3 \sum_{j=1}^{N_r} \Big) \, w_j^3 \ \delta_{\ell 0}
 \label{SNbisp2thin}
\eeqa

%
%

In computing the bispectrum, an additional complexity over the power spectrum case is that a search over closed triangles has to be done~\cite{ScoColFry9803},  enforced by the delta function in Eqs.~(\ref{BispMono}) and~(\ref{BispQuad}). There are many ways to do this, let us briefly discuss two options that we have found reasonably efficient and implemented in the past~\cite{Sco00,FelFriFry0102,ScoFelFry0101}. One is to use a fast ($N \ln N$)  algorithm such as {\sf quicksort} to sort Fourier coefficients into shells to quickly find $\q_3=-\q_{12}$ in shell $k_3$ given $\q_1$ in shell $k_1$ and $\q_2$ in shell $k_2$. The sorting  can be done once and the results stored in disk when multiple realizations need to be run (e.g. when running on mock catalogs) since it only depends on grid and bin size.  The other is to use FFTs themselves to find closed  triangles by using the plane-wave representation of the delta function and factorizing the estimator in real space, i.e. 
\beqa
&& \prod_{i=1}^3 \, \int_{k_i} {d^3q_i}\, \dD(\q_{123}) \, F_{\ell_1}(\q_1) F_{\ell_2}(\q_2) F_{\ell_3}(\q_3) \nonumber \\
&&= \int {d^3 x \over (2\pi)^3} \ F_{k_1}^{(\ell_1)}(\x)\, F_{k_2}^{(\ell_2)}(\x)\, F_{k_3}^{(\ell_3)}(\x)
\label{Factorize}
\eeqa
where
\beq
F_{k}^{(\ell)}(\x) \equiv \int_k  d^3q\ {\rm e}^{i \q\cdot \x} \, F_\ell(\q)
\label{F_k}
\eeq
and thus for each bin $k_i$ one must do an inverse FFT to find the $F_{k}^{(\ell)}$, and then sum over real space to obtain the bispectrum for a given $k_1,k_2,k_3$. This estimator can be trivially extended to higher-order spectra, as the trispectrum~\cite{Sef0512}.

Figure~\ref{Qell} shows the results of measuring the bispectrum for $\ell=0,2,4$ for the LasDamas LRG ($M_g<-21.8$) DR7 mock catalogs for triangles that correspond to $k_1=0.047 \kMpc$, $k_2=2\, k_1$ as a function of the angle $\theta$ between $\k_1$ and $\k_2$. The bispectrum multipoles were computed using Eq.~(\ref{BispQuad}) with the shot noise correction given by Eq.~(\ref{SNbisp2thin}). We see that the typical  dependence on triangle shape of the reduced bispectrum defined as 
\beq
Q_{123}^{(\ell)}={B_{123}^{(\ell)}\over P_0(k_1)P_0(k_2)+P_0(k_2)P_0(k_3)+P_0(k_3)P_0(k_1)}
\label{Qelldef}
\eeq
is shared among the three multipoles. This was already known up to $\ell=2$ in the plane-parallel approximation~\cite{ScoCouFri9906}. We leave the comparison of these measurements to theoretical predictions for an upcoming paper.

\subsection{Performance}

To get an idea of performance of the estimators presented here, we now discuss for definiteness run times of our codes for the northern part of the CMASS sample in the Baryon Oscillation Spectroscopic Survey~\cite{AndAubBai1406}, which comprises about $N_g = 650,000$ galaxies (and $N_r \approx 100 N_g$). All timings in what follows are for a single processing core. The galaxies and random objects are put into a cartesian box of $3.6 \Gpc$ a side and fourth-order interpolated into interlaced $360^3$ grids, enough to reach $k=0.3\kMpc$ for the power spectrum and $k=0.2\kMpc$ for the bispectrum. To give an idea of how far we are from the plane-parallel approximation, for the DR11 mask we have 
\beqa
& & 
L_{xx}\simeq 0.50, \ \ \
L_{yy} \simeq 0.275, \ \ \ 
L_{zz} \simeq 0.222, \ \ \  \nonumber \\ 
& & 
L_{xy}\simeq 0.02, \ \ \
L_{yz} \simeq -0.025, \ \ \ 
L_{zx} \simeq -0.24, \ \ \ 
\label{Qat0}
\eeqa
where 
\beq
L^{ab} \equiv { \sum_j^{N_r} \hat{x}_j^a \hat{x}_j^b\, w_j \, {\rm e}^{i\k\cdot \x_j} \over  \sum_j^{N_r}  w_j \, {\rm e}^{i\k\cdot \x_j}}
\label{Lab}
\eeq
which shows dominance of the $x$-direction but significant amplitudes in the other directions as well.

The galaxies and randoms are interpolated and the 7 FFTs computed. For the galaxies this takes 22 seconds, while for the randoms 210 seconds. Computing the additional 15 FFTs (and interpolations) as well takes about 60 seconds in total for the galaxies, while computing the corresponding FFTs for $F_\ell^w(\q)$ (needed for bispectrum shot noise subtraction) doubles the timings again (for a total of 44 interpolations and FFTs).  The overdensity fields are constructed and the $\ell=0,2,4$ multipoles from $k_{\rm min}=0.0052 \kMpc$ up to $k_{\rm max}=0.3 \kMpc$ in $\delta k = 0.0052 \kMpc$ bins computed in 2 seconds. For the bispectrum we compute monopole and quadrupole for all triangles with sides between $k_{\rm min}=0.0052 \kMpc$ and $k_{\rm max}=0.209 \kMpc$ in $\delta k = 0.0052 \kMpc$ bins in about 11 minutes; adding the hexadecapole yields an additional 5 minutes. Clearly, this is a more than adequate performance and can be scaled to significantly larger data sets with no foreseeable issues (in fact we have routinely used these algorithms for tens of billion particle simulations in the plane-parallel case). In addition, most importantly, this performance allows us to estimate the covariance matrix of these estimators using $\sim 10^4$ survey realizations with realistic masks and noise properties, as has been already presented in the past for smaller surveys~\cite{Sco00,2006PhRvD..74b3522S}.

%

\section{Conclusions}
\label{conclude}

Computing the redshift-space clustering of galaxies is the key goal of large-scale structure and one of the most sensitive probes to  test gravity and dark energy, bias and primordial non-Gaussianity. Application to wide surveys demands defining statistics that characterize the effect of redshift-space distortions allowing for the angular dependence of the line of sight along which distortions operate. 

In this paper, starting from a definition of local estimators that generalize the definition of power spectrum and bispectrum to the case of lack of statistical homogeneity, we defined natural estimators for power spectrum and bispectrum multipoles. In the case of the power spectrum multipoles, our natural estimator agrees for $\ell=2$ with~\cite{YamNakKam0602}, while for the bispectrum multipoles it is new. We then considered slight modifications to these estimators in which the ``center of mass line of sight" is allowed to rotate among the different members of a pair (for the power spectrum) or triplet (for the bispectrum). As a result of this, we presented very efficient estimators to calculate the power spectrum and bispectrum multipoles, which require only FFTs. 

For the power spectrum quadrupole, an additional 6 FFTs are required over the monopole, while for the power spectrum hexadecapole we presented two estimators, one that requires an additional 15 FFTs over the quadrupole, and another one which does not require any additional FFTs. We implemented both in mock catalogs with realistic geometries and found that while the two hexadecapole estimators agree with each other in the mean value, the more expensive estimator has a somewhat lower cosmic variance, and thus higher signal to noise. We also showed first results for the bispectrum $\ell=0,2,4$ estimators in mock catalogs and  presented specifics about the performance of such estimators for the BOSS survey. Their speed and scaling makes application to larger future datasets such as eBOSS, Euclid and DESI quite encouraging. In the near future we will present detailed studies of the bias of these estimators when compared to the plane-parallel approximation (which is always used to make theoretical predictions) as well as their application to the DR12 sample of BOSS galaxies.

\vskip 2pc

When this paper was in preparation~\cite{BiaGilRug1505} appeared on the arXiv with the same results  for the FFT factorization of the power spectrum multipoles estimator leading to our Eqs.~(\ref{delta2b}) and~(\ref{delta4b}).  

\acknowledgements

I thank Mart\'{\i}n Crocce, Di Liu, Ariel S\'anchez, Emiliano Sefusatti, and Jeremy Tinker for discussions. This work was supported by NSF grant AST-1109432. 
 I thank the CERN theory group for a sabbatical visit during Fall 2014 where most of this work was done. 

\bibliography{masterbiblio}


\end{document}